\documentclass[preprint,showpacs,preprintnumbers,amsmath,amssymb]{revtex4}

\usepackage{graphicx}
\usepackage{dcolumn}
\usepackage{bm}

\begin{document}

\preprint{Submission to Phys. Rev. B}

\title{
Negative magnetization of Li$_2$Ni$_2$Mo$_3$O$_{12}$ having 
a spin system composed of distorted honeycomb lattices and linear chains
}

\author{Masashi Hase$^1$}
 \email{HASE.Masashi@nims.go.jp}
\author{Vladimir Yu. Pomjakushin$^2$}
\author{Vadim Sikolenko$^3$}
\author{Lukas Keller$^2$}
\author{Hubertus Luetkens$^4$}
\author{Andreas D\"onni$^1$}
\author{Hideaki Kitazawa$^1$}

\affiliation{%
${}^{1}$National Institute for Materials Science (NIMS), 1-2-1 Sengen, Tsukuba, 
Ibaraki 305-0047, Japan \\
${}^{2}$Laboratory for Neutron Scattering, Paul Scherrer Institut (PSI), 
CH-5232 Villigen PSI, Switzerland \\
${}^{3}$Karlsruhe Institute of Technology, 
Kaiserstrasse 12, 76131 Karlsruhe, Germany \\
${}^{4}$Laboratory for Muon-Spin Spectroscopy, Paul Scherrer Institut (PSI), 
CH-5232 Villigen PSI, Switzerland
}%

\date{\today}

\begin{abstract}

We studied magnetism of a spin-1 substance Li$_2$Ni$_2$Mo$_3$O$_{12}$. 
The spin system consists of distorted honeycomb lattices and linear chains of Ni$^{2+}$ spins. 
Li$^+$ ions enter about 25 \% and 50 \% of honeycomb and chain Ni sites, respectively, 
creating disorder in both the spin subsystems. 
A magnetic phase transition occurs at $T_{\rm c} = 8.0$ K in the zero magnetic field. 
In low magnetic fields, the magnetization 
increases rapidly below $T_{\rm c}$, 
decreases below 7 K, and 
finally becomes negative at low temperatures. 
We determined the magnetic structure using neutron powder diffraction results. 
The honeycomb lattices and linear chains show 
antiferromagnetic and ferromagnetic long-range order, respectively. 
We investigated static and dynamic magnetic properties 
using the local probe technique muon spin relaxation.
We discuss the origin of the negative magnetization. 

\end{abstract}

\pacs{75.25.-j, 75.30.Cr, 75.40.Cx, 75.47.Lx, 76.75.+i}

\maketitle

\section{INTRODUCTION}

Several antiferromagnets having plural magnetic ions or magnetic-ion sites 
show an interesting magnetic long-range order (LRO). 
In Cu$_2$Fe$_2$Ge$_4$O$_{13}$, 
Cu$^{2+}$ spin dimers (spin 1/2) are coupled to 
Fe$^{3+}$ spin chains (spin 5/2).\cite{Masuda04,Masuda05}
An indirect Fe-Fe exchange coupling via Cu dimers was observed.\cite{Masuda07}
This result reveals that 
the Cu dimers play the role of nonmagnetic media in the indirect magnetic interaction.
Dynamics of dimer excitation in staggered and random fields generated by Fe spins 
are also experimentally identified.\cite{Masuda09}
In Cu$_3$Mo$_2$O$_9$, 
three crystallographically independent Cu$^{2+}$ sites exist (Cu1, Cu2, and Cu3). 
Spins (1/2) on Cu1 sites form antiferromagnetic (AFM) chains.\cite{Hamasaki08} 
Two spins on neighboring Cu2 and Cu3 sites form an AFM dimer. 
Each AFM chain is coupled to AFM dimers. 
Only a component parallel to the $b$ axis of magnetic moments on Cu1 sites 
is ordered below the AFM transition temperature $T_{\rm N} = 7.9$ K 
in the zero magnetic field.\cite{Hamasaki08}
Perpendicular components are ordered in magnetic fields parallel to the $a$ or $c$ axis.
A canted AFM LRO is stabilized only in finite magnetic fields.  
It is inferred that magnetic competition causes this unique LRO. 
In addition, the canted AFM LRO disappears in Cu$_3$Mo$_2$O$_9$ 
with a very small amount doping ({\it e.g.,} 0.2 \% Zn for Cu sites).\cite{Hase08}

Negative magnetization is an interesting phenomena. 
It was first observed in Li$_{0.5}$(FeCr)$_{2.5}$O$_4$ spinels \cite{Gorter53} and 
was explained using the N\'eel model.\cite{Neel48}
In this model, 
two collinear sublattice magnetizations ($M_{\rm A}$ and $M_{\rm B}$) 
with different temperature ($T$) dependence 
are considered. 
$|M_{\rm A}|$ is larger than $|M_{\rm B}|$ below  $T_{\rm N}$, while 
$|M_{\rm B}|$ is larger than $|M_{\rm A}|$ below a compensation temperature $T_{\rm comp}$. 
If directions of the sublattice magnetizations are fixed, 
magnetization is observed as a negative value below $T_{\rm comp}$.   
Several compounds showing negative magnetization have been found. 
Some of them are considered to be explainable by the 
N\'eel model,\cite{Menyuk60,Yasukochi60,Schulkes63,Luthi66,Abe71,Zakaria05,Kimishima00,Kimishima05,Ivanov03,Hashimoto05} 
while for the other compounds the origin is different or has not been determined 
explicitly.\cite{Cooke74,Washimiya78,Yoshii01a,Yoshii00,Yoshii01b,Khomchenko08,Andrzejewski06,Shirakawa91,Mahajan92,Nguyen95,Ren98,Ren00,Blake02,Adachi99,Singh08a,Singh08b,Singh10a,Singh10b,Duman02} 
Details of the origins will be described later.  

We have paid our attention to Li$_2$Ni$_2$Mo$_3$O$_{12}$. 
Two crystallographically independent Ni$^{2+}$ (spin 1) sites exist ({\it M}1 and {\it M}2) 
as represented in Fig. 1.\cite{Ozima77} 
The {\it M}1 and {\it M}2 sites form distorted honeycomb lattices and linear chains, respectively. 
About 25 \% of {\it M}1 sites and about 50 \% of {\it M}2 sites 
are occupied by non-magnetic Li$^{+}$ ions 
as shown by the formula Li(Li$_{0.5}$Ni$_{1.5}$)(Li$_{0.5}$Ni$_{0.5}$)Mo$_3$O$_{12}$. 
Unexpectedly, we found negative magnetization at low $T$. 
To investigate the origin of the negative magnetization, 
we determined the magnetic structure of Li$_2$Ni$_2$Mo$_3$O$_{12}$ 
using the neutron powder diffraction technique and 
determined its static and dynamic magnetic properties 
using the local probe technique muon spin relaxation ($\mu$SR).

\begin{figure}
\begin{center}
\includegraphics[width=8cm]{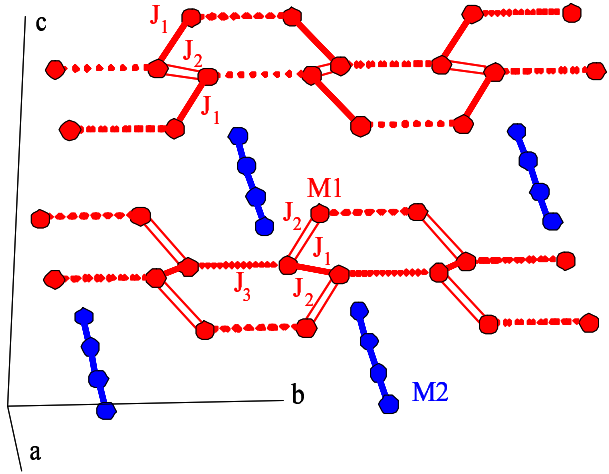}
\caption{
(Color online)
A schematic drawing of {\it M}1 ($8d$) and {\it M}2 ($4c$) sites in Li$_2$Ni$_2$Mo$_3$O$_{12}$. 
The space group is orthorhombic $Pnma$ (No. 62).\cite{Ozima77} 
These sites are occupied by 
Ni$^{2+}$ (spin 1) or Li$^+$ (spin 0) ions. 
The occupation ratio of Ni$^{2+}$ ions is 0.757 at {\it M}1 and 0.566 at {\it M}2. 
The ratio between Ni numbers at {\it M}1 and {\it M}2 sites is 0.728 : 0.272. 
Three short {\it M}1-{\it M}1 bonds exist and form distorted honeycomb lattices. 
The {\it M}1-{\it M}1 distances are 3.03, 3.08, and 3.75 \AA \ at 10 K for 
the first, second, and third shortest {\it M}1-{\it M}1 bonds, respectively. 
The respective exchange interaction parameters are defined as 
$J_1$, $J_2$, and $J_3$. 
The shortest {\it M}2-{\it M}2 bonds form linear chains. 
The {\it M}2-{\it M}2 distance is 2.53 \AA. 
Distances in the other ${\it M}i - {\it M}i$ bonds ($i = 1$ or 2) are larger than 5.06 \AA. 
Distances in {\it M}1-{\it M}2 bonds are larger than 5.19 \AA. 
}
\end{center}
\end{figure}

\section{Experimental Methods}

Crystalline powder of Li$_2$Ni$_2$Mo$_3$O$_{12}$ 
was synthesized using a solid-state-reaction method 
at 973 K in air for 144 h with intermediate grindings. 
We used an isotope $^{7}$Li (purity: 99 \%) for neutron-diffraction samples 
to decrease absorption of neutrons. 
We confirmed formation of Li$_2$Ni$_2$Mo$_3$O$_{12}$ 
using an X-ray diffractometer (JDX-3500; JEOL).
Li$_2$Ni$_2$Mo$_3$O$_{12}$ is insulating at room temperature. 

We measured magnetization 
using a superconducting quantum interference device 
(SQUID) magnetometer (MPMS-5S; Quantum Design). 
We entered powder in paraffin molten by heating in the zero magnetic field and 
fixed finally the powder in solid paraffin. 
We measured magnetization of the powder embedded in solid paraffin. 
Probably, powder directions are randomly distributed and 
powder reorientation does not occur in the magnetization measurements. 
Specific heat was measured using relaxation technique with 
Physical Property Measurement System (PPMS; Quantum Design). 
We used a sintered pellet in the specific-heat measurements. 

We determined the magnetic structure of 
$^7$Li$_2$Ni$_2$Mo$_3$O$_{12}$ 
from neutron powder diffraction data. 
The experiments were conducted
using the high-resolution powder diffractometer for thermal neutrons 
HRPT \cite{hrpt}
(wavelength $\lambda = 1.886$ \AA) and 
the high-intensity cold neutron powder diffractometer
DMC ($\lambda = 4.206$ \AA) 
at the Swiss spallation neutron source SINQ in PSI. 
Powder was filled into a vanadium container 
with 8 mm diameter and 55 mm height. 
Rietveld refinements of diffraction data were performed 
using the {\tt FULLPROF Suite} program package.\cite{Rodriguez93} 
Symmetry analyses of possible magnetic configurations 
were conducted using the program {\tt BASIREP} 
in the  {\tt FULLPROF Suite} program package. 

The muon spin relaxation ($\mu$SR) measurements were performed 
using the Dolly spectrometer at the $\pi$E1 beamline 
at the Swiss Muon Source (S$\mu$S) in PSI. 
In a $\mu$SR experiment nearly 100~\% spin-polarized muons are
implanted into the sample one at a time. The positively charged
$\mu^+$ thermalize at interstitial lattice sites, where they act
as magnetic micro probes. In a magnetic material the muon spin
precesses about the local magnetic field $B$ at the muon site with
the Larmor frequency $f_{\mu} = \gamma_\mu/(2\pi) B$ (muon
gyromagnetic ratio $\gamma_\mu /2 \pi = 135.5$ MHz~T$^{-1}$). With
a lifetime of $\tau_\mu = 2.2$~$\mu$s the muon decays into two
neutrinos and a positron, the latter being predominantly emitted
along the direction of the muon spin at the moment of the decay.
Measurement of both the direction of positron emission as well as
the time between muon implantation and positron detection for an
ensemble of several millions of muons provides the time evolution
of the muon spin polarization $P(t)$ along the initial muon spin
direction. In a powder, 2/3 of the local magnetic field components
are perpendicular to the $\mu^+$ spin and cause a precession,
while the 1/3 longitudinal field components do not. The damping of
the oscillation is a measure of the width of the static field
distribution experienced by the muon ensemble. In a static
magnetic environment, the 1/3 fraction of the muon spin
polarization is conserved. In a dynamic magnetic materials however
also this so-called 1/3-tail relaxes and the relaxation rate can
be, in certain limits, related to magnetic fluctuation rates.
Since $\mu$SR is a local probe technique, the amplitudes of the
different signals observed are a measure of the corresponding
volume fractions. With this, it provides a direct measure of
magnetic volumes which is not easily possible with scattering
techniques. For further details of the $\mu$SR technique, we refer
to the recent review in Ref.~\cite{Yaouanc}.

\section{Experimental Results}

Figure 2 portrays the $T$ dependence of 
the specific heat in the zero magnetic field. 
A peak is apparent at 8.0 K. 
As it will be shown later, magnetization increases rapidly on cooling below 8.0 K. 
Therefore, the peak indicates occurrence of 
a magnetic phase transition at $T_{\rm c} = 8.0$ K. 

\begin{figure}
\begin{center}
\includegraphics[width=8cm]{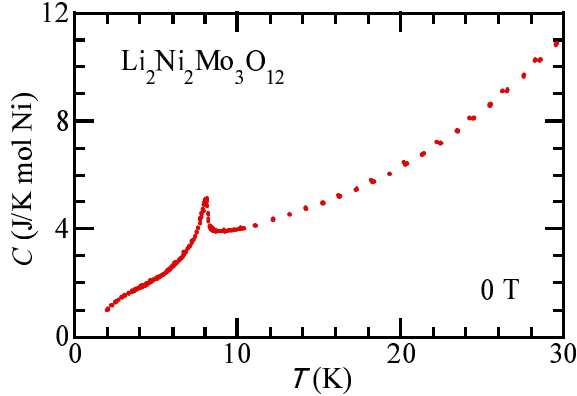}
\caption{
(Color online)
Temperature dependence of the specific heat of Li$_2$Ni$_2$Mo$_3$O$_{12}$ 
in the zero magnetic field. 
}
\end{center}
\end{figure}

Figure 3 (a) shows the $T$ dependence of 
the magnetization $M$ of Li$_2$Ni$_2$Mo$_3$O$_{12}$ 
in a very weak residual magnetic field. 
The value of the magnetic field was estimated as 
$6.3 \times 10^{-5}$ \ T (0.63 Oe) 
from paramagnetic magnetization values in this field and 0.1 T. 
The magnetization increases rapidly below 8.2 K, 
shows a maximum around 7.2 K, and 
decreases below 7.2 K. 
It should be noted that the magnetization is negative below 4.2 K. 
We did not observe any difference between magnetizations 
measured for increasing and decreasing $T$. 
Figure 4 represents the magnetic field $H$ dependence of $M$ at 1.7 K 
after the residual field cooling process.
We can see a small residual magnetization and hysteresis. 
The rapid increase below 8.2 K and small residual magnetization indicate
appearance of a canted AFM LRO or ferromagnetic (FM) LRO with small magnetic moments.  
Figure 3 (b) shows the $T$ dependence of $M$  in the field of $1 \times 10^{-3}$ \ T. 
We can see a hysteresis between magnetizations 
measured in the  residual field cooling (RFC) and 
field cooling (FC) processes. 
The absolute value of $M$ in the FC process is larger than 
that in the RFC process.  
In 0.1 T [Fig. 3(d)], only the FC magnetization becomes negative. 
In 0.2 T [Fig. 3(e)], both the RFC and FC magnetizations are positive. 
In 1 T [Fig. 3(f)], the hysteresis is small. 
Figure 3 (g) shows the $T$ dependence of 
paramagnetic susceptibility in 0.1 T. 
The susceptibility above 20 K obeys the Curie-Weiss law. 
We determined the value of the Curie constant as 1.12 emu K/mol Ni from the data above 200 K. 
The $g$ value was calculated as 2.11 on the assumption that the spin value is 1. 
This $g$ value is reasonable for Ni$^{2+}$ spins. 

\begin{figure}
\begin{center}
\includegraphics[width=8cm]{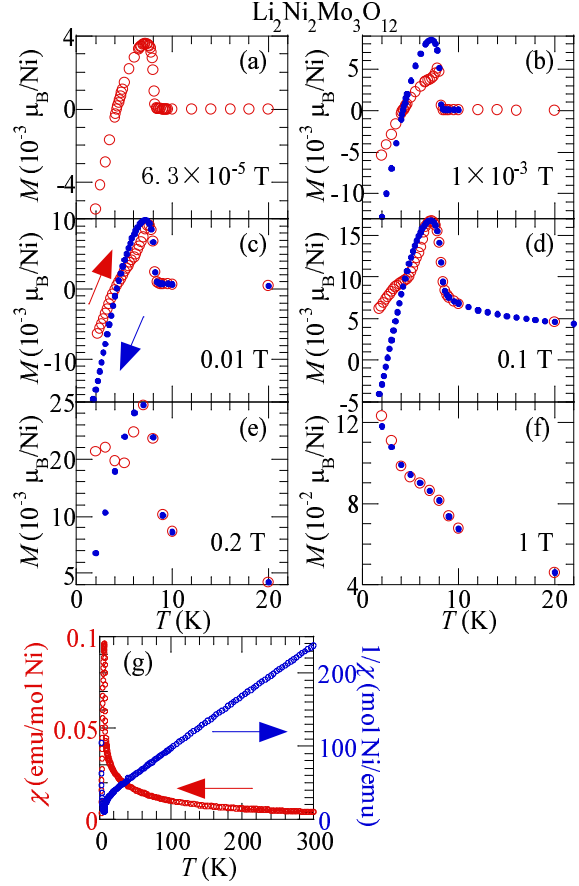}
\caption{
(Color online)
(a) - (f)
Temperature dependence of the magnetization of Li$_2$Ni$_2$Mo$_3$O$_{12}$ 
at low $T$ in various magnetic fields. 
Open and closed circles indicate data measured in the RFC and FC processes, respectively.
(g)
Temperature dependence of the magnetic susceptibility and its inverse 
in the paramagnetic state. 
}
\end{center}
\end{figure}

\begin{figure}
\begin{center}
\includegraphics[width=8cm]{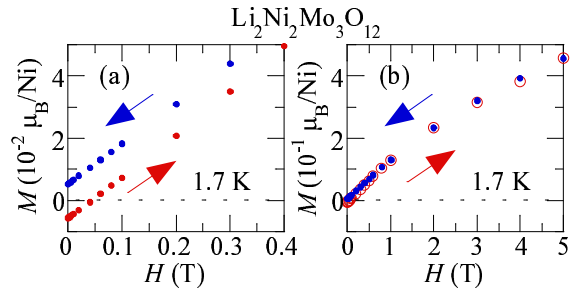}
\caption{
(Color online)
Magnetic field dependence of the magnetization of Li$_2$Ni$_2$Mo$_3$O$_{12}$ 
at 1.7 K. 
Arrows indicate directions of magnetic field scans. 
}
\end{center}
\end{figure}

Figure 5 depicts the neutron powder diffraction pattern of 
paramagnetic $^7$Li$_2$Ni$_2$Mo$_3$O$_{12}$ recorded 
using the HRPT diffractometer 
with $\lambda = 1.886$ \AA \ at 10 K, 
which is slightly higher than
$T_{\rm c} = 8.0$ K.
The refinement based on the crystal structure of 
Li$_2$Ni$_2$Mo$_3$O$_{12}$ as determined by room temperature 
single-crystal X-ray diffraction \cite{Ozima77}  
well fits the experimental neutron diffraction pattern at 10 K.
Structural parameters are presented in Table I.  

\begin{figure}
\begin{center}
\includegraphics[width=8cm]{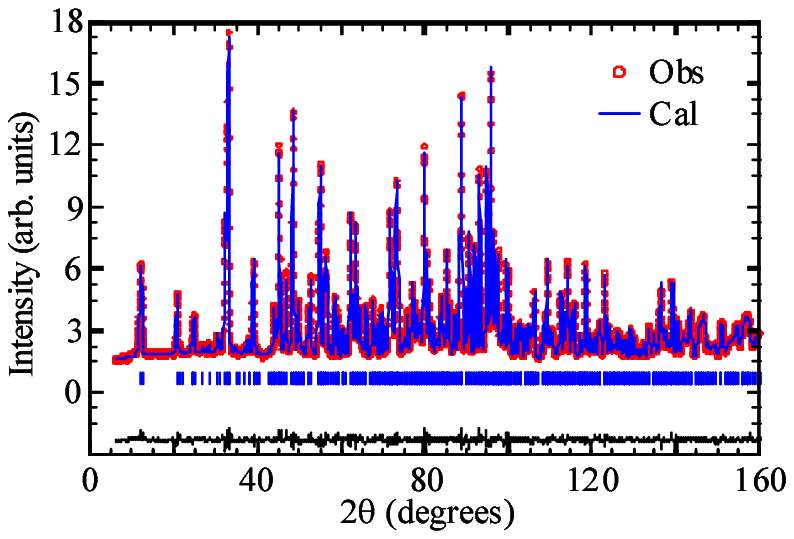}
\caption{
(Color online)
A neutron powder diffraction pattern of $^7$Li$_2$Ni$_2$Mo$_3$O$_{12}$ 
at 10 K (higher than $T_{\rm c}$) measured using the HRPT diffractometer ($\lambda = 1.886$ \AA). 
Lines on the observed pattern and at the bottom 
show a Rietveld refined pattern and difference between the observed and 
the Rietveld refined patterns. 
Hash marks represent positions of nuclear reflections.
}
\end{center}
\end{figure}

\begin{table*}
\caption{\label{table1}
Structural parameters of $^7$Li$_2$Ni$_2$Mo$_3$O$_{12}$ derived from 
Rietveld refinement of
the HRPT neutron powder diffraction pattern at 10 K.
The space group is orthorhombic $Pnma$ (No. 62).  
The lattice constants at 10 K are
$a=5.0639(5)$ \AA, $b=10.397(1)$ \AA, and $c=17.467(2)$ \AA. 
Estimated standard deviations are shown in parentheses. 
The atomic displacement parameters $B_{\rm iso}$ were constrained to be the same for 
the same atom types and for the atoms occupying the same site. 
The reliability factors of the refinement amounted to 
$R_{\rm wp}=3.21$, $R_{\rm exp}=1.77$, $\chi^2=3.31$, and $R_{\rm Bragg}=2.76$
}
\begin{ruledtabular}
\begin{tabular}{ccccccc}
Atom & Site & $x$ & $y$ & $z$ & $B_{\rm iso}$ \AA$^2$  & Occupancy\\
\hline
Ni1 & 8{\it d} & 0.7473(3) & 0.4302(1) & 0.0258(1) & 0.27(3) & 0.757(2)\\
Li1 & 8{\it d} & 0.7473(3) & 0.4302(1) & 0.0258(1) & 0.27(3) & 0.243(2)\\
Ni2 & 4{\it c} & 0.3881(6) & 0.75 & 0.2497(2) & 0.27(3) & 0.566(2)\\ 
Li2 & 4{\it c} & 0.3881(6) & 0.75 & 0.2497(2) & 0.27(3) & 0.434(2)\\ 
Li3 & 4{\it c} & 0.7562(13) & 0.25 & 0.1967(3) & 0.27(3) &\\
Mo1 & 4{\it c} & 0.7149(4) & 0.75 & 0.0562(1) & 0.06(2) &\\     
Mo2 & 8{\it d} & 0.2225(3) & 0.4739(1) & 0.1562(1) & 0.06(2) &\\       
 O1 & 4{\it c} & 0.5657(5) & 0.25 & 0.0064(1) & 0.29(2) &\\       
 O2 & 4{\it c} & 0.6353(5) & 0.75 & 0.1544(1) & 0.29(2) &\\       
 O3 & 8{\it d} & 0.1383(4) & 0.6184(2) & 0.2050(1) & 0.29(2) &\\       
 O4 & 8{\it d} & 0.4233(4) & 0.3779(2) & 0.2137(1) & 0.29(2) &\\       
 O5 & 8{\it d} & 0.9411(4) & 0.3864(2) & 0.1256(1) & 0.29(2) &\\       
 O6 & 8{\it d} & 0.4214(4) & 0.5122(2) & 0.0740(1) & 0.29(2) &\\  
 O7 & 8{\it d} & 0.9163(4) & 0.6146(2) & 0.0363(1) & 0.29(2) &\\        
\end{tabular}
\end{ruledtabular}
\end{table*}

Figure 6(a) shows 
difference between two neutron powder diffraction patterns of 
$^7$Li$_2$Ni$_2$Mo$_3$O$_{12}$ at 2.0 K and 10 K 
collected using the DMC diffractometer ($\lambda = 4.206$ \AA). 
The reflections
of the difference pattern can be indexed in the chemical cell with the
propagation vector ${\bf k} = 0$. 
The profile matching le Bail fit, in
which the integral intensities are the refined parameters, 
shows that all the peaks are well described with {\bf k} = 0
with the reliability factors $R_{\rm wp}=2.34$, $R_{\rm exp}=1.72$, and 
$\chi^2=1.86$. 
The new reflections appear below $T_{\rm c}$. 
Therefore, they must be magnetic reflections. 

Using the determined propagation
vector, we performed a symmetry analysis according to Izyumov {\it et
al.}\cite{Izyumov91} to derive possible magnetic configurations for
the space group $Pnma$. 
The space group has eight one dimensional irreducible representations (IRs). 
The observed magnetic patterns were
compared with calculated patterns using the structural parameters
determined from the structural refinement. 
After sorting out the basis functions of all eight IRs, 
we found that only $\tau_5$ well fits the experimental pattern.  
The IR $\tau_5$ has the following characters 1, -1, 1, -1,
1, -1, 1, -1  for the symmetry elements listed in the caption of Fig.~7. 
The magnetic moments of Ni2 atoms are restricted by $\tau_5$ symmetry
to be parallel to $b$-axis, thus forming a ferromagnetic sublattice.
The  $y$-components of Ni1 spins are also forced to be ferromagnetically aligned,
whereas the spins in $xz$-plane have antiferromagnetic configuration
with basis vectors listed in the caption of Fig.~7. 

\begin{figure}
\begin{center}
\includegraphics[width=8cm]{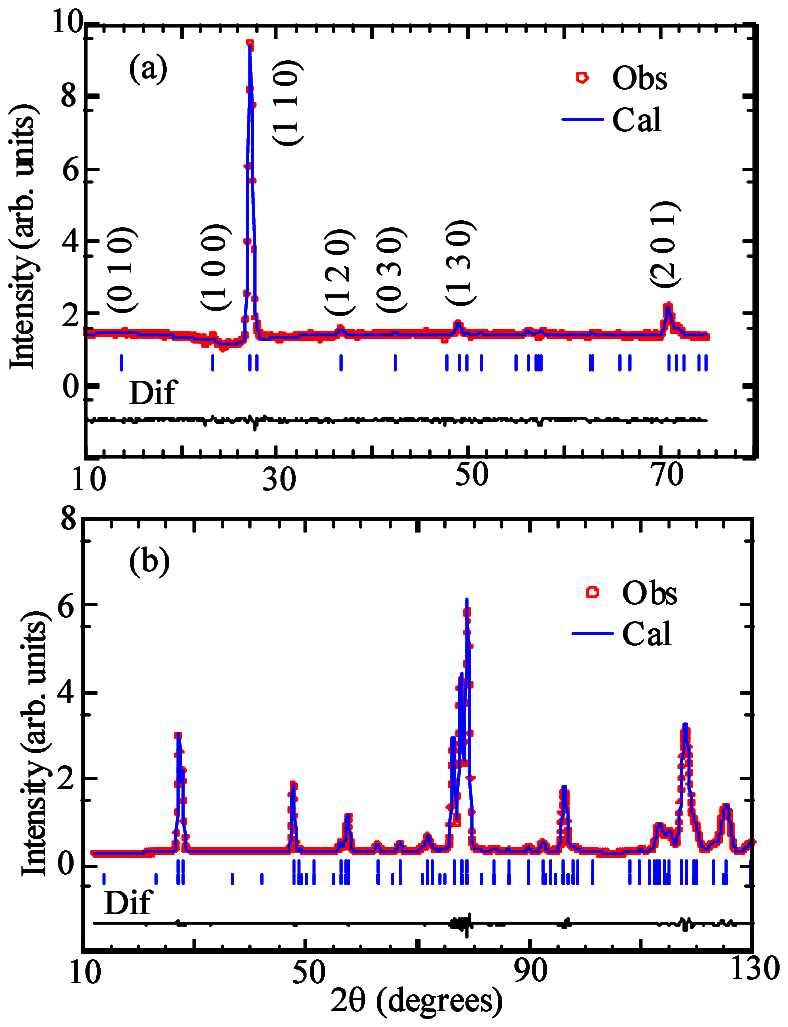}
\caption{
(Color online)
(a)
A fragment of the difference between two neutron powder diffraction patterns of $^7$Li$_2$Ni$_2$Mo$_3$O$_{12}$ 
at 2.0 K and 10 K (magnetic neutron diffraction pattern) 
measured using the DMC diffractometer ($\lambda = 4.206$ \AA)
Lines on the observed pattern and at the bottom 
show a Rietveld refined pattern in le Bail profile matching mode 
and difference between the observed and the Rietveld refined patterns. 
Hash marks represent positions of magnetic reflections. 
Several indices of the reflections are written. 
(b)
A neutron powder-diffraction pattern at 2.0 K.  
Lines on the observed pattern and at the bottom 
show a Rietveld refined pattern for the magnetic structure model $\tau_5$
and difference between the observed and the Rietveld refined patterns. 
Upper and lower hash marks represent positions of nuclear and magnetic reflections, respectively. 
}
\end{center}
\end{figure}

In the refinement of the magnetic structure, 
all the atom structure parameters were fixed by 
the values obtained from HRPT data (Table I). 
The magnetic structure is depicted in Fig. 7.
The magnetic moment at the Ni1(1) position is 
(0.27(4), 0.04(12), -1.58(3))$\mu_{\rm B}/{\rm Ni}$ and its magnitude is 1.60(3) $\mu_{\rm B}/{\rm Ni}$ at 2.0 K .  
The ordered Ni1 moment mainly points along the $c$ direction 
with a small component along the $a$ direction. 
The size of the ferromagnetic $b$ component is refined to a small value close to zero
within experimental errorbars. 
The arrangement of the moments indicates that 
the $J_1$, $J_2$, and $J_3$ interactions are 
FM, FM, and AFM, respectively. 
The magnetic moment at the Ni2(1) position is 
(0, -0.6(1), 0)$\mu_{\rm B}/{\rm Ni}$. 
Ni2 moments form the FM LRO. 
The Ni2 moment is reduced in comparison with 
ordinary ordered Ni$^{2+}$ moments. 
The reduction may be caused by the one dimensionality of the Ni2 spin system and 
large disorder due to about  50 \ \% substitution of Li ions. 

\begin{figure}
\begin{center}
\includegraphics[width=8cm]{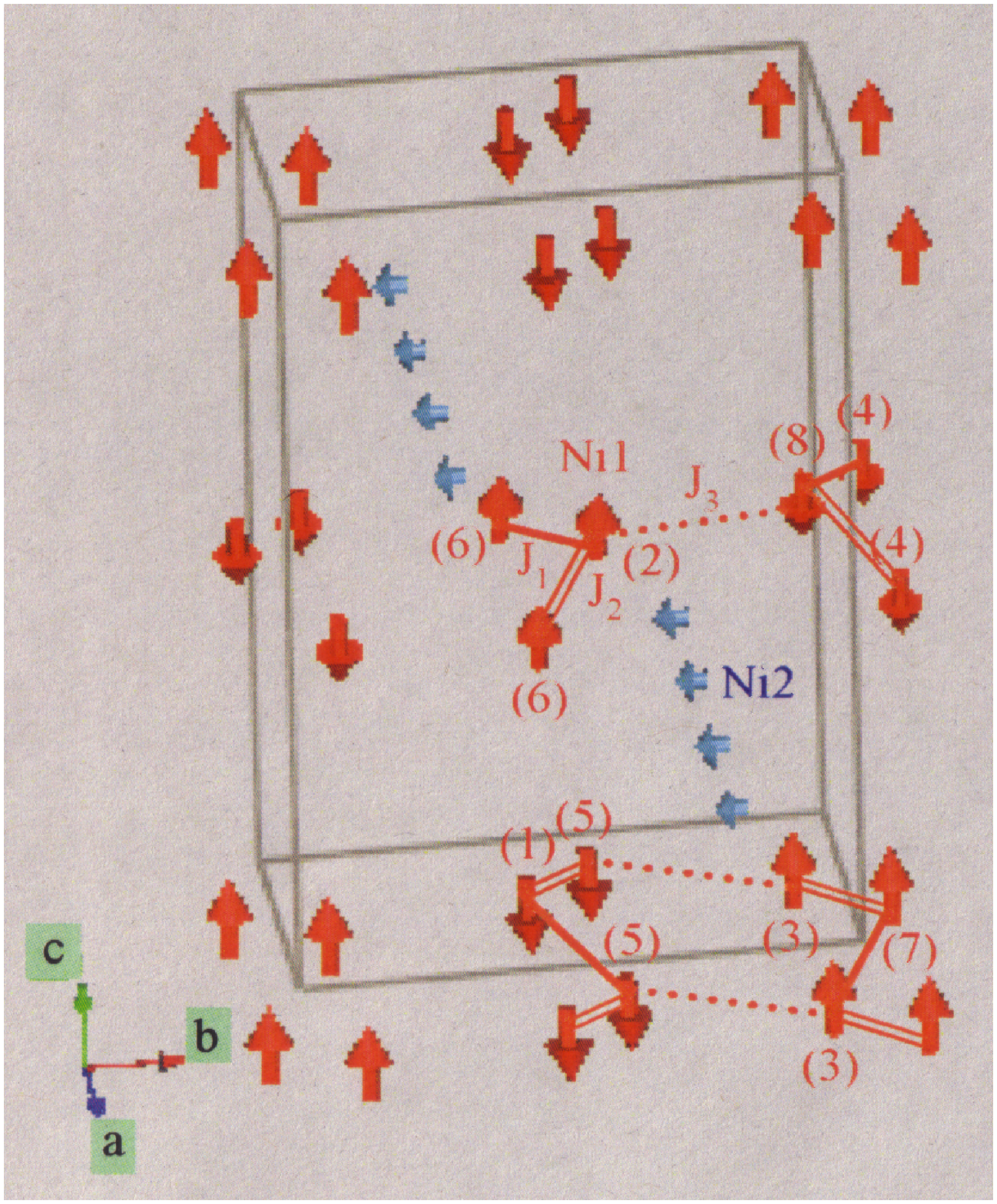}
\caption{
(Color online)
The magnetic structure of Li$_2$Ni$_2$Mo$_3$O$_{12}$. 
Ni2 moments form FM LRO. 
To show clearly the arrangement of Ni1 moments, 
symmetry operators of {\it M}1 (8d) sites are given.  
(1) $x, y, z [u, v, w]$,  
(2) $\bar x +1/2, \bar y, z + 1/2 [u, v, \bar w]$, 
(3) $\bar x, y + 1/2, \bar z [\bar u, v, \bar w]$, 
(4) $x + 1/2, \bar y + 1/2, \bar z + 1/2 [\bar u, v, w]$, 
(5) $\bar x, \bar y, \bar z [u, v, w]$, 
(6) $x +1/2, y, \bar z + 1/2 [u, v, \bar w]$, 
(7) $x, \bar y + 1/2, z [\bar u, v, \bar w]$, and 
(8) $\bar x + 1/2, y + 1/2, z + 1/2 [\bar u, v, w]$. 
Here $u$, $v$, and $w$ indicate components of the magnetic moment. 
}
\end{center}
\end{figure}

\begin{figure}
\begin{center}
\includegraphics[width=8cm]{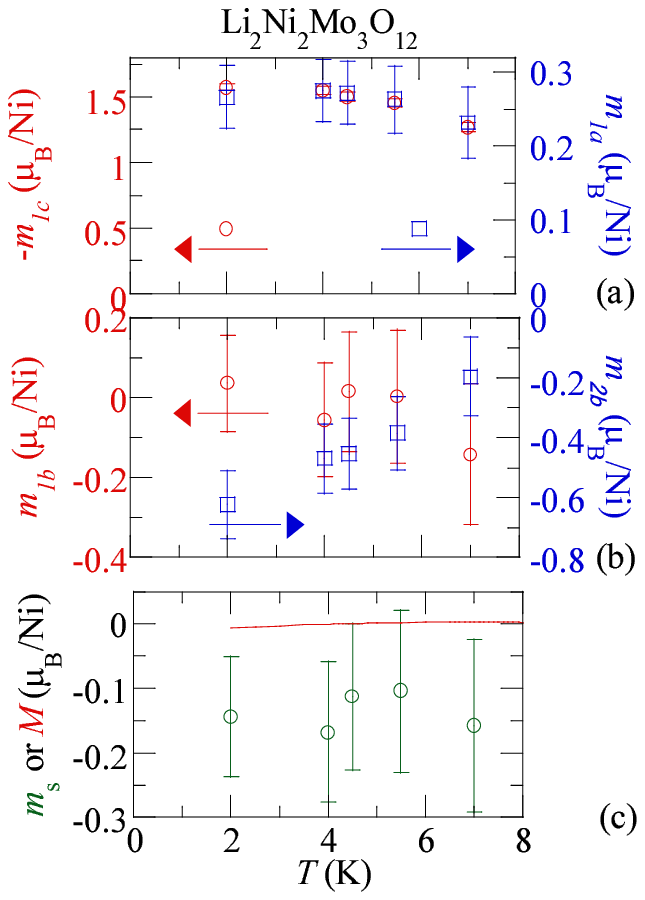}
\caption{
(Color online)
Temperature dependence of the components of ordered magnetic moments 
in Li$_2$Ni$_2$Mo$_3$O$_{12}$.  
(a)
$-m_{1c}$ and $m_{1a}$.
(b)
$m_{1b}$ and $m_{2b}$.  
(c)
The average spontaneous magnetization per Ni $m_{\rm s}$. 
The solid line represents the RFC magnetization shown in Fig. 3(a). 
}
\end{center}
\end{figure}

The IR $\tau_5$ well fits experimental data for all the measured temperatures. 
Figure 8 shows the $T$ dependence of the magnetic moments. 
The absolute value of the FM component $m_{2b}$ increases monotonically with decreasing $T$. 
The AFM components $m_{1a}$ and $m_{1c}$ have significantly weaker dependence
of temperature with step like change at $T_{\rm c}$. 
The FM component $m_{1b}$ stays close to zero with slight tendency to increase.   

\begin{figure}
\begin{center}
\includegraphics[width=8cm]{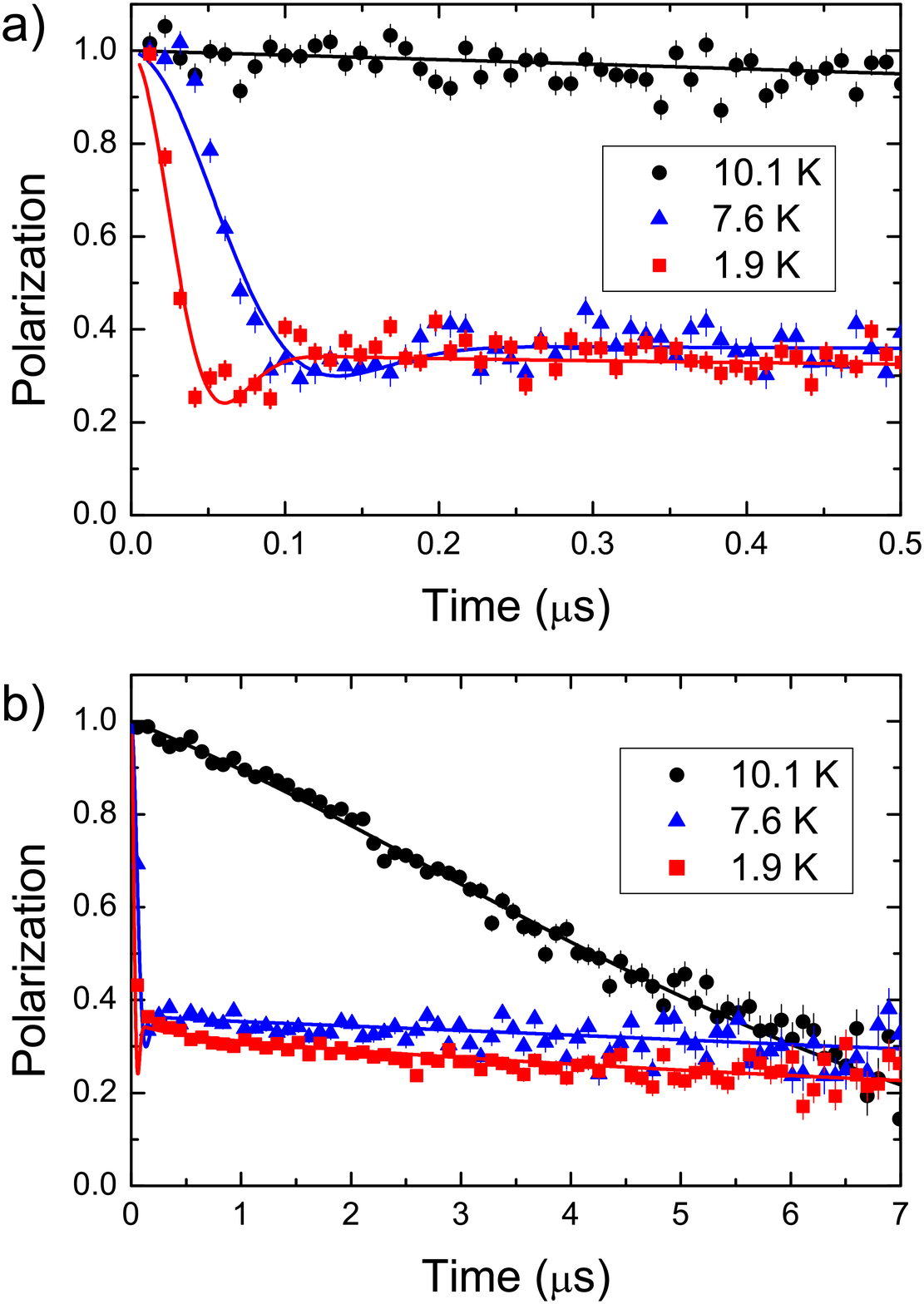}
\caption{
Zero field muon spin relaxation spectra for various
characteristic temperatures shown on a short (a) or a long (b) time
scale.} \label{muSR-Spectra}
\end{center}
\end{figure}

To further elucidate the static and dynamic magnetic properties of Li$_2$Ni$_2$Mo$_3$O$_{12}$, 
we performed zero field $\mu$SR experiments for various temperatures. 
In Fig. \ref{muSR-Spectra}
zero field $\mu$SR spectra are shown for characteristic
temperature above and below $T_{\rm c} = 8.0$ K. 
Figures \ref{muSR-Spectra} (a) and (b) show the same data but
plotted on different time scales to highlight the observed changes
for the fast and slow relaxing signals. At high temperatures a
slowly relaxing spectrum of Gauss-Kubo-Toyabe form \cite{kubo} is
observed. This is the fingerprint of a non-magnetic material. The
small Gaussian relaxation $\sigma_{nm}$ is due to static random
nuclear moments only. 
Below $T_{\rm c} = 8.0$ K 
an over-damped spontaneous muon spin precession is observed, 
indicating a very broad field distribution at the muon site(s), 
i.e. a complicated or disordered magnetic structure or many muon sites 
in the crystallographic lattice. 
This result is consistent with structural disorder caused by Li ions. 

The $\mu$SR spectra have been fitted
using the following functional form for the polarization function:
\begin{equation}
P(t) = (1 - f_{mag})\,G_{nm}(t) + f_{mag}\,G_{mag}(t)
\end{equation}
with
\begin{equation}
G_{nm}(t) =
\left(\frac{2}{3}\,[1-(\sigma_{nm}\,t)^2]\,e^{-\frac{1}{2}(\sigma_{nm}\,t)^2}
+ \frac{1}{3} \right)\,e^{-\lambda_{dyn}\,t}
\end{equation}
and
\begin{equation}
G_{mag}(t) = \left(\frac{2}{3}\,\cos(2\pi f_\mu
t)\,e^{-\frac{1}{2}(\sigma_{mag}\,t)^2}+\frac{1}{3}\right)\,e^{-\sqrt{\lambda_{dyn}\,t}}.
\label{root}
\end{equation}
Here $G_{mag}$ and $G_{nm}$ represent the relaxation functions for
the magnetic ($f_{mag}$) and non-magnetic ($1-f_{mag}$) volume
fractions with their corresponding relaxation rates $\sigma_{mag}$
and $\sigma_{nm}$, respectively. $\lambda_{dyn}$ is the dynamic
relaxation rate.

\begin{figure}
\begin{center}
\includegraphics[width=8cm]{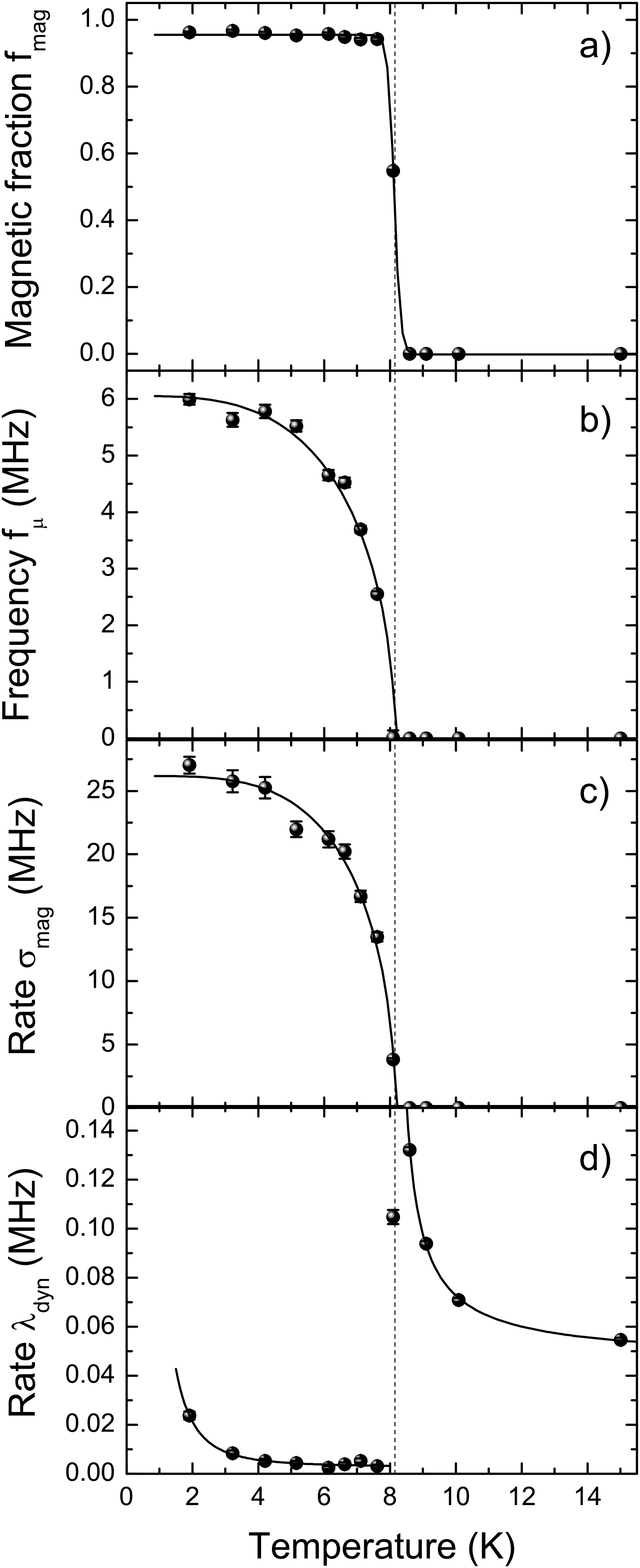}
\caption{
Parameter values obtained from the fits to
 the zero field muon spin relaxation measurements. 
The lines are guides to the eye.
} \label{muSR-para}
\end{center}
\end{figure}

Figure~\ref{muSR-para} shows the obtained parameter values from the fits
using the above equations.
A sharp magnetic transition at $T_{\rm c} = 8.0$~K is observed in all the parameters. 
The magnetic fraction $f_{mag}$ reaches
$\approx$~100~\% shortly below $T_{\rm c}$. The magnetic order parameter
measured via the $\mu$SR frequency $f_\mu$ as well as the static
field width $\sigma_{mag}$ continuously increase below $T_{\rm c}$.
The values of $\sigma_{mag}$ are larger than the values of $f_\mu$, 
indicating the very broad width of the internal field distribution.  
Approaching the transition from above an additional exponential
damping which increases towards $T_{\rm c}$ is observed. This is typical
for the slowing down of electronic moments approaching $T_{\rm c}$. The
corresponding dynamical relaxation rate $\lambda_{dyn}$ is shown
in Fig.~\ref{muSR-para}~(d). 
Below the transition $\lambda_{dyn}$
is nearly zero indicating a static magnetic state. Decreasing the
temperature further, a dynamic relaxation of the 1/3 tail is
observed again. This is already clearly visibly from the
comparison of the $\mu$SR spectra at 7.6~K and 1.9~K in
Fig.~\ref{muSR-Spectra}~(b). The dynamic relaxation at low
temperature had to be fitted with a root-exponential time
dependence (Eq.~\ref{root}), which is indicative of a
broad distribution of fluctuation times. 
Interestingly, the dynamic relaxation rate starts to increase below $\approx 4$~K 
where the magnetization becomes negative. 

\section{Discussion}

Let us consider the origin of the negative magnetization in Li$_2$Ni$_2$Mo$_3$O$_{12}$. 
The N\'eel model requires two FM sublattices that in our case are formed
by the $m_{1b}$ and $m_{2b}$ components. 
The values of $m_{2b}$ are negative. 
The values of $m_{1b}$ are very small and show both the signs. 
At 2.0 K, the error bar of $m_{1b}$ is the smallest and 
the value of $m_{1b}$ is positive. 
Therefore, 
if the $T$ dependence of $m_{1b}$ is similar to that of $-m_{1c}$ or $m_{1a}$, 
$m_{1b}$ may be intrinsically positive. 
Compensation of $m_{1b}$ and $m_{2b}$ is possible. 
The sublattice magnetizations $M_{\rm A}$ and $M_{\rm B}$ defined in the Introduction correspond to 
$m_{1b}$ and $m_{2b}$, respectively.  
$M_{\rm A}$ has a weaker $T$ dependence than $M_{\rm B}$ in the N\'eel model. 
If $m_{1b}$ is proportional to $-m_{1c}$ or $m_{1a}$, 
$m_{1b}$ has a weaker $T$ dependence than $m_{2b}$ as the N\'eel model. 
We calculated the average spontaneous magnetization per Ni $m_{\rm s} = p_1 m_{1b} + (1-p_1)m_{2b}$ and 
plotted it together with the RFC magnetization in Fig. 8(c). 
Here, the proportion coefficient $p_1=0.728$ is calculated from the occupancies listed in Table I.  
The contribution of $m_{1b}$ is large in $m_{\rm s}$
because of the large number of Ni1 atoms. 
Although the component $m_{2b}$ shows well defined decrease with temperature lowering [Fig. 8(b)], 
$m_{\rm s}$ is practically constant [Fig. 8(c)].  
The difference in the $T$ dependence of $m_{2b}$ and $m_{\rm s}$ 
suggests compensation of the FM components. 
A reversal of $m_{\rm s}$, however, was not obtained.  
The absolute values of $m_{\rm s}$ seem much larger than 
the absolute values of the RFC magnetization represented by the solid line in Fig. 8(c).  

The negative magnetization is too small to be detected 
using neutron powder diffraction techniques. 
Therefore, we could not verify the validity of the N\'eel model. 
The magnetic scattering length was calculated as $2.7 g S f(Q)$ fm using a Born approximation. 
Here, $g$ and $S$ are the $g$ value and magnitude of ordered spins, respectively. 
$f(Q)$ is the magnetic form factor and its magnitude is in the order of unity or less. 
The nuclear coherent scattering length of Ni was obtained as 10.3 fm (natural average).\cite{NeutronDataBooklet}
The intensity of Bragg peaks is proportional to the square of scattering lengths. 
The lower limit of magnetic moments that can be evaluated in neutron powder diffraction experiments  
is about 0.1 $\mu_{\rm B}$. 
As is seen in Fig. 8, error bars are in the order of 0.1 $\mu_{\rm B}$. 

The following four alternative models may be considered as the candidates 
for explanation of the negative magnetization effect.
In model 1 applicable to GdCrO$_3$ or (La{\it RE})CrO$_3$ 
({\it RE} = Pr or Nd),\cite{Cooke74,Washimiya78,Yoshii01a,Yoshii00,Yoshii01b,Khomchenko08} 
magnetic moments of Cr ions form a canted AFM LRO, while 
{\it RE} ions are paramagnetic.  
A Dzyaloshinskii-Moriya (DM) interaction exists between Cr and {\it RE} spins. 
The main ($x$) component of the Cr moment 
produces an internal magnetic field parallel to the $z$ direction on {\it RE} ions. 
The polarization of paramagnetic moments on {\it RE} ions induced by the internal magnetic field 
is aligned opposite to the canted components parallel to the $z$ direction. 
Magnitude of sum of the canted components is larger than 
magnitude of sum of the polarization of the paramagnetic magnetization 
below $T_{\rm N}$, while 
the latter becomes larger than the former below $T_{\rm comp}$. 
The direction of the main component of the Cr moments is fixed. 
Therefore, the direction of the canted component is also fixed. 
As a result, the negative magnetization appears in 
GdCrO$_3$ or (La{\it RE})CrO$_3$ ({\it RE} = Pr or Nd).
In Li$_2$Ni$_2$Mo$_3$O$_{12}$, 
the distance between Ni1 and Ni2 sites is larger than  5.19 \AA. 
Probably, exchange interactions between Ni1 and Ni2 spins are weak. 
Internal magnetic fields produced by Ni1 (Ni2) moments on Ni2 (Ni1) sites 
are small.   
Accordingly, the negative magnetization in Li$_2$Ni$_2$Mo$_3$O$_{12}$ 
cannot be explained by the model 1.

In model 2 applicable to LaVO$_3$ \cite{Shirakawa91,Mahajan92,Nguyen95}
or YVO$_3$,\cite{Ren98,Ren00,Blake02} 
only the canted component of the magnetic moment on V ions reverses and 
becomes anti-parallel to external fields. 
Therefore, the negative magnetization was observed. 
In  LaVO$_3$, a structural phase transition occurs just below $T_{\rm N}$ 
due to the Jahn-Teller effect. 
As a result, the DM vector reverses and therefore 
the canted component of the magnetic moment reverses. 
In YVO$_3$, 
the canting direction is determined mainly by 
the single-ion anisotropy at high $T$ and by the DM interaction at low $T$. 
As a result, the canting direction reverses at low $T$. 
In Li$_2$Ni$_2$Mo$_3$O$_{12}$, 
LRO of Ni2 spins is the FM order (not canted AFM order). 
Consequently, the model 2 cannot account for 
the negative magnetization in Li$_2$Ni$_2$Mo$_3$O$_{12}$. 

In model 3 applicable to (SmGd)Al$_2$,\cite{Adachi99}  
compensation between the spin and the orbital parts of the ordered moments occurs. 
We observe a sum of spin and orbital magnetic moments. 
Therefore, the results in Fig. 8 indicate directly 
that compensation between the spin and the orbital parts of the ordered moments is not the origin of the negative magnetization. 
In addition, in 3d shells, only spin magnetic moments exist because of quenching of the orbital angular momentum. 

In model 4 possibly applicable to Sr$_2${\it RE}RuO$_6$ 
({\it RE}= Yb, Y or Lu),\cite{Singh08a,Singh08b,Singh10a,Singh10b} 
exchange bias (EB) effects \cite{Nogues99} may cause negative magnetization. 
The effects can appear in composites having both FM and AFM LROs. 
Ordered magnetic moments of the two LROs should be collinear. 
The FM transition temperature should be higher than the AFM transition temperature. 
In Li$_2$Ni$_2$Mo$_3$O$_{12}$, however,  
Ni1 and Ni2 moments are nearly perpendicular to each other. 
Only one transition was observed within experimental accuracy. 
In addition, 
distances in Ni1 - Ni2 bonds are larger than 5.19 \AA, suggesting that 
interactions between Ni1 and Ni2 spins are weak. 
Consequently, EB effects cannot account for the negative magnetization in Li$_2$Ni$_2$Mo$_3$O$_{12}$. 

At present, 
we think that the N\'eel model may be valid for the negative magnetization in Li$_2$Ni$_2$Mo$_3$O$_{12}$, 
although we cannot prove the validity of  the N\'eel model. 
We cannot deny perfectly possibility of another (unknown) origin. 
As was described in the $\mu$SR results, 
the dynamic relaxation rate starts to increase below $\approx 4$~K 
where the magnetization becomes negative. 
It is tempting to conclude that 
the magnetization inversion mechanism involves a dynamical process. 
As another idea, a structural change, which is not detected, may cause reversal of $m_{2b}$. 
Experimental investigations into Li$_2$Ni$_2$Mo$_3$O$_{12}$ using other techniques and 
further theoretical considerations are necessary. 

\section{Summary}

We studied magnetism of the spin-1 substance Li$_2$Ni$_2$Mo$_3$O$_{12}$. 
The spin system is composed of the distorted honeycomb lattices and linear chains 
formed by two distinct Ni$^{2+}$ sites, Ni1 and Ni2 sites, respectively. 
Li$^+$ ions substitute about 25 \% and 50 \% of the honeycomb and chain Ni sites, respectively, 
creating the disorder in both the spin subsystems. 
A magnetic phase transition occurs at $T_{\rm c} = 8.0$ K. 
In low magnetic fields, the magnetization 
increases rapidly just below $T_{\rm c}$, 
decreases below 7 K, and 
finally becomes negative at low temperatures on cooling. 
We determined the magnetic structure using neutron powder diffraction results. 
The magnetic order on both Ni-sites develops 
according to a single irreducible representation $\tau_5$ of the space group $Pnma$ and propagation vector ${\bf k}=0$.  
The honeycomb lattices show antiferromagnetic (AFM) long-range order. 
The ordered Ni1 moment mainly points along the $c$ direction 
with a small AFM component along the $a$ direction. 
The size of the ferromagnetic (FM) $b$ component is very small. 
The ordered Ni2 moment on linear chains has only the $b$ component and 
shows FM long-range order. 
The magnetic moment sizes amounted to 
1.60(3)and 0.6(1) $\mu_B$ per Ni-atom 
for the honeycomb-lattice and chain Ni sites, respectively. 
The sharp magnetic transition at $T_{\rm c}$ was also observed in the $\mu$SR results. 
The over-damped spontaneous muon spin precession below $T_{\rm c}$ 
indicates a very broad field distribution at the muon site(s). 
The dynamic relaxation rate starts to increase below $\approx 4$~K 
where the magnetization becomes negative. 
We have examined several models known from literature that can account for the negative magnetization. 
The N\'eel model that is based on the compensation of the ferromagnetic sublattices 
may be valid in our case, but our experimental data does not allow us to prove this model quantitatively. 

\begin{acknowledgments}

We are grateful 
to H. Mamiya for fruitful discussion, 
to S. Matsumoto for sample syntheses and X-ray diffraction measurements, and 
to M. Kaise for X-ray diffraction measurements.  
The neutron powder diffraction experiments were conducted
at SINQ, PSI Villigen, Switzerland. 
The $\mu$SR measurements were performed 
at the Swiss Muon Source, PSI Villigen, Switzerland.
This work was partially supported by grants from NIMS. 

\end{acknowledgments}

\newpage 

\end{document}